\documentclass[fleqn,usenatbib,useAMS]{mnras}

\usepackage{newtxtext,newtxmath}
\usepackage{physics}
\usepackage[T1]{fontenc}

\DeclareRobustCommand{\VAN}[3]{#2}
\let\VANthebibliography\thebibliography
\def\thebibliography{\DeclareRobustCommand{\VAN}[3]{##3}\VANthebibliography}

\usepackage{graphicx}	
\usepackage{amsmath}	

\title[VLBA CANDELS GOODS-North Survey - I]{The VLBA CANDELS GOODS-North Survey. I \-- Survey Design, Processing, Data Products, and Source Counts }

\author[R.P. Deane et al.]{Roger P. Deane$^{1,2}$\thanks{E-mail: roger.deane@wits.ac.za},
Jack F. Radcliffe$^{2,3}$,
Ann Njeri$^{3,4}$,
Alexander Akoto-Danso$^{5,6}$,
Gianni Bernardi$^{7,5,8}$,
\newauthor
Oleg M. Smirnov$^{5,8}$,
Rob Beswick$^{3}$,
Michael A. Garrett$^{3,9}$,
Matt J. Jarvis$^{10,11}$, 
Imogen H. Whittam$^{10,11}$,
\newauthor
Stephen Bourke$^{12,13}$,
Zsolt Paragi$^{14}$
\\
$^{1}$Wits Centre for Astrophysics, School of Physics, University of the Witwatersrand, 1 Jan Smuts Avenue, Johannesburg, 2000, South Africa \\
$^{2}$Department of Physics, University of Pretoria, Private Bag X20, Pretoria 0028, South Africa\\
$^{3}$Jodrell  Bank  Centre  for  Astrophysics,  School  of  Physics  \&  Astronomy,  The  University  of  Manchester,  Alan  Turing  Building, Oxford Road, Manchester M13 9PL, UK\\
$^{4}$School of Mathematics, Statistics \& Physics, Newcastle University, NE1 7RU, Newcastle Upon Tyne, UK\\
$^{5}$Department of Physics and Electronics, Rhodes University, PO Box 94, Makhanda 6140, Eastern Cape, South Africa\\
$^{6}$Ghana Space Science and Technology Institute, Ghana Atomic Energy Commission, Proton Street, Legon, Accra, Ghana \\
$^{7}$INAF-Istituto di Radio Astronomia, via Gobetti 101, 40129 Bologna, Italy \\
$^{8}$South African Radio Astronomy Observatory, 2 Fir Street, Observatory, 7925, South Africa\\
$^{9}$Leiden Observatory, Leiden University, PO Box 9513, 2300 RA Leiden, The Netherlands \\
$^{10}$Sub-Department of Astrophysics, University of Oxford, Keble Road, Oxford OX1 3RH, UK \\
$^{11}$Department of Physics and Astronomy, University of the Western Cape, Robert Sobukwe Road, 7535 Bellville, Cape Town, South Africa \\
$^{12}$  Department of Space, Earth and Environment, Chalmers University of Technology, Onsala Space Observatory, 439 92 Onsala,
Sweden \\
$^{13}$Overstock Ireland Ltd, Westgate, Finisklin Business Park, Sligo, Ireland F91 HF66 \\
$^{14}$Joint Institute for VLBI in Europe, Oude Hoogeveensedijk 4, NL-7991 PD, Dwingeloo, The Netherlands \\
}

\date{Accepted 2024 January 18. Received 2024 January 18; in original form 2023 November 6}

\pubyear{2022}

\begin{document}
\label{firstpage}
\pagerange{\pageref{firstpage}--\pageref{lastpage}}
\maketitle

\begin{abstract}
The past decade has seen significant advances in wide-field cm-wave very long baseline interferometry (VLBI), which is timely given the wide-area, synoptic survey-driven strategy of major facilities across the electromagnetic spectrum. While wide-field VLBI poses significant post-processing challenges that can severely curtail its potential scientific yield, many developments in the km-scale connected-element interferometer sphere are directly applicable to addressing these. Here we present the design, processing, data products, and source counts from a deep (11 $\mu$Jy\,beam$^{-1}$), quasi-uniform sensitivity, contiguous wide-field (160 arcmin$^2$) 1.6 GHz VLBI survey of the CANDELS GOODS-North field. This is one of the best-studied extragalactic fields at milli-arcsecond resolution and, therefore, is well-suited as a comparative study for our Tera-pixel VLBI image. The derived VLBI source counts show consistency with those measured in the COSMOS field, which broadly traces the AGN population detected in arcsecond-scale radio surveys. However, there is a distinctive flattening in the $ S_{\rm 1.4GHz}\sim$100-500\,$\mu$Jy flux density range, which suggests a transition in the population of compact faint radio sources, qualitatively consistent with the excess source counts at 15 GHz that is argued to be an unmodelled population of radio cores. This survey approach will assist in deriving robust VLBI source counts and broadening the discovery space for future wide-field VLBI surveys, including VLBI with the Square Kilometre Array, which will include new large field-of-view antennas on the African continent at $\gtrsim$1000~km baselines. In addition, it may be useful in the design of both monitoring and/or rapidly triggered VLBI transient programmes.
\end{abstract}

\begin{keywords}
surveys -- galaxies: active -- techniques: interferometric -- techniques: high angular resolution
\end{keywords}



\section{Introduction} \label{sec:intro}

This decade sees a slew of cm-wavelength, arcsec-scale radio surveys with wide areal coverage ($\gtrsim$1~deg$^2$). The primary science objectives thereof include measuring the cosmic star formation rate history; constraining the population of low-luminosity and Compton-thick active galactic nuclei (AGN); improved statistical understanding of jet triggering and mechanical feedback; the relation to neutral hydrogen and molecular gas reservoirs in galaxies and large-scale structure; image-plane transient and variability searches; as well as a range of cosmological applications through weak lensing and multi-wavelength cross-correlation experiments enabled by wide and deep surveys at these angular scales \citep[e.g.][and references therein]{Schinnerer2010,Norris2011,Condon2012,Heywood2016,Jarvis2016,Smolcic2017,Murphy2017,Heywood2020,Muxlow2020,Chowdhury2022,Heywood2022,Hurley-Walker2022,Best2023}. These arcsecond-scale interferometric surveys at cm-wavelengths cover a wide range in depth and area on facilities including MeerKAT, the upgraded Karl G. Jansky Very Large Array, the Australian SKA Pathfinder Telescope, the Giant Metrewave Radio Telescope, the Murchison Widefield Array, and the International LOFAR Telescope. This survey-driven paradigm is motivated by statistical requirements of the above-mentioned scientific objectives; however, they typically lack higher angular resolution counterparts to address several long-standing questions faced in galaxy and AGN evolution, including the relative radio emission contributions from star formation and AGN-related activity for spatially unresolved, low-luminosity sources, as well as the discovery potential for scientifically-rich individual sources (e.g. binary supermassive black holes, gravitational lenses). In this pursuit, Very Long Baseline Interferometry (VLBI) can play a pivotal role as it spatially filters high brightness temperature emission and isolates radio cores and compact jets, hence constraining the compact AGN contribution. The enormous strides made over the past decade in VLBI survey area and depth make this a compelling approach, in concert with multi-wavelength programmes, to survey supermassive black hole accretion in the Universe if processing constraints can be overcome. 

Since the first pioneering steps in wide-field VLBI \citep{Garrett1996,Garrett1999,Garrett2001,Lenc2008,Chi2013} it has been clear that the post-processing was an impracticably expensive computational task to perform using typical approaches, even for moderate fractions ($\ll 0.1$) of the available field-of-view. The problem stems from the required time and frequency resolution required to avoid significant sensitivity losses due to time and bandwidth smearing \cite[e.g.][]{TMS2017}. The required resolution to achieve this scales with baseline length, meaning that VLBI arrays require orders of magnitude higher time and frequency resolution to process the full field-of-view when compared to a km-scale interferometer. This results in many orders of magnitude larger data rates and storage requirements despite significantly lower source sky densities at a given flux density threshold. This has not hampered progress in VLBI since its first half-century has been almost exclusively focused on pointed observations of single objects at the centre of the narrow processed field-of-view.  The VLBI visibility data are thus heavily averaged in both time and frequency since the processed and imaging field-of-view is typically restricted to only regions of at most a few arcseconds away from the pointing centres or phase centres. Given the low cm-VLBI sky source density at a $\gg 1$~mJy sensitivity level, it was relatively seldom that there would be multiple detectable sources in the field before major bandwidth upgrades, so processed fields-of-view were typically limited to $\lesssim 1$~arcsec$^{2}$, rather than attempting to image the available field-of-view on the order of $\sim$0.1-1\,deg$^{2}$ \citep[with notable exceptions, of course, e.g.][]{Garrett1996,Chi2013}. 

In order to image a non-negligible fraction of the available field-of-view while minimizing point-source-sensitivity loss, the required time and frequency resolution leads to large processing demands, which are not scalable to wide-area surveys ($\gg1$~deg$^2$) with the current archiving and correlator capacity of VLBI network operators. A transformational step in wide-field VLBI capability was the multi-phase centre correlation technique \citep{Morgan2011}, which was implemented in the DiFX software correlator \citep{Deller2011} and the SFXC correlator \citep{Keimpema2015}. This approach saves orders of magnitude in data volume over full-field imaging and has the added computational benefit of parallelised data processing streams. This has enabled wide-field VLBI observations of sources with a flux density of a few tens of $\mu$Jy with relative ease and enabling more efficient wide-field VLBI surveys \citep[e.g.][]{Middelberg2011,Middelberg2013,Deller2014,Herrera2017,Radcliffe2018,Petrov2021}. Contemporary wide-field VLBI surveys therefore have two possible strategies: (i) record the data at sufficiently high time and frequency resolution in order to image the entire region of interest, or (ii) image a number of considerably smaller sub-regions centred on a catalogue of phase centres, the positions of which are typically selected from known arcsec-scale radio or multi-wavelength detections.

Using the latter approach, recent wide-field VLBI surveys have demonstrated that VLBI is an integral and unique tool in the statistical study of AGN activity over cosmic time. The current state-of-the-art for deep VLBI extragalactic fields is the COSMOS VLBA Survey \citep{Herrera2017} which employed the multi-phase centre technique to detect 468 VLA-detected sources down to an average noise rms $\sigma \sim 10 \, \mu$Jy\,beam$^{-1}$ in the 2\,deg$^2$ COSMOS field. This was followed up with a $\sim$3$\times$ deeper, narrower tier which included the Greenbank Telescope and identified 35 sources below the `VLBA-only' sensitivity, enabling the deepest VLBI source counts constraints to date. Of particular interest and utility in these surveys is a more detailed understanding of how radio source counts at milliarcsecond scales compared to the more readily available, higher brightness temperature sensitivity arcsec-scale radio surveys. This is a key step toward a statistical determination of what dominates the contribution of radio flux for different source populations.  Therefore, there is a strong motivation for improved statistical power in the number of VLBI detections at $\lesssim 10~\mu$Jy flux density levels in legacy multi-wavelength extragalactic fields, enabling robust host galaxy characterisation and an enriched astrophysical analysis.

In this paper, we present a wide-field VLBI survey that aims to build on the technical progress described above and enhance our statistical understanding of compact radio sources. In Section~\ref{sec:surveyscience} and Section~\ref{sec:surveydesign}, we outline the motivation and design of a quasi-uniform VLBI survey of an extragalactic legacy field. In Section~\ref{sec:obs}, \ref{sec:cal}, and \ref{sec:imaging}, we describe the technical details of the observations, calibration, imaging and source-finding techniques required to achieve this. Section~\ref{sec:dataproducts} presents the cross-calibration catalogue and images. In Section~\ref{sec:counts}, we derive the differential radio source counts from this uniform area survey and compare with them the COSMOS field VLBI source counts \citep{Herrera2018}, alongside arcsec-scale radio surveys \citep[e.g.][]{Smolcic2017,Hale2023}.

The data products from this survey will be used to carry out several analyses presented later in this paper series. These include a detailed comparison with other radio surveys, host galaxy properties, and analysis of the origin of the radio emission in Njeri et al.~(in press, Paper II hereafter); a 12-epoch VLBI transient and variability search over $>$2 months; as well as leveraging the uniform sensitivity and VLBA's homogeneity to carry out a systematic study of statistical self-calibration schemes \citep[e.g.][]{Middelberg2013,Radcliffe2016}.

\section{Survey Science and Technical Drivers}\label{sec:surveyscience}

Here, we outline the primary motivations for carrying out a quasi-uniform sensitivity survey over a deep, multi-wavelength extragalactic legacy field, as compared to the more traditional approach of selected phase centres on known radio detections from arcsec-scale interferometers. As discussed, different aspects of these will be presented in independent papers in the series, however, we briefly summarize them below. 

\begin{enumerate}
    \item Serendipitous discovery of variable/transient sources (e.g. AGN, low-redshift radio supernovae) not identified or present in previous radio observations \citep[e.g.][]{Bower2007,Stewart2016,Perley2017,Radcliffe2019}. 
    \item The measurement of differential VLBI source counts at $\sim$10\,$\mu$Jy\,beam$^{-1}$ sensitivity over a well-defined area for direct comparison with arcsec-scale radio surveys to better understand the relevant source populations. 
    \item Statistical approaches to the scientific analysis enabled by the deep multi-wavelength coverage, including high-resolution imaging from the \emph{Hubble Space Telescope (HST)}, Atacama Large Millimeter/submillimeter Array (ALMA), and the \emph{James Webb Space Telescope (JWST)} in the future \citep[e.g.][]{Lindroos2016,Inami2020}. 
    \item A systematic study of statistical self-calibration  (i.e. multi-source self-calibration, \citealt{Middelberg2013,Radcliffe2016}) using a homogeneous VLBI array, comparing the tradeoff between using a large number of marginally-detected sources with a small number of higher signal-to-noise detections. 
    \item Scientifically useful lower limits on the compact, high-brightness radio emission within all galaxies catalogued within the selected extragalactic field, which can assist multi-wavelength AGN classification if sufficiently deep \citep[e.g.][]{Whittam2022}. 
\end{enumerate}

Maximising the above required careful selection of the target extragalactic legacy field for this quasi-uniform VLBI survey experiment, which naturally has an AGN and galaxy evolution focus. A key requirement was the need for deep, high-resolution multi-wavelength coverage from radio through X-ray. Furthermore, previous VLBI observations were desirable as a comparison of known sources identified and characterized using the traditional VLBI approach \citet{Chi2013}. We selected the CANDELS GOODS-North field as optimal for the above purpose. In addition, the field's Declination of $+62$~deg results in a favourable \emph{uv}-coverage with the VLBA, resulting in a point spread function with comparatively low sidelobes and, therefore, relatively high imaging fidelity. We describe some of the key science drivers in surveying this field and extensions to it below. These are addressed in the current paper and subsequent papers in the series.\\
\\
\noindent{\bf Comparison of low-luminosity AGN source counts from milli-arcsecond to arcsecond scales: }
The unambiguous identification of radio-quiet, low-luminosity AGN remains a challenge with arcsec-resolution radio surveys, particularly as they reach $\mu$Jy-level sensitivity \citep[see, e.g.][and references therein]{Padovani2016}. These increasingly sensitive and wide area surveys show broad consistency but also some deviations from model predictions \citep[e.g.][]{Wilman2008,Bonaldi2019}. For example, \citet{Whittam2013,Whittam2017,Whittam2020} use 15~GHz radio observations to show that there are a population of faint radio sources which are not included in simulated source counts \citep[e.g.][]{Wilman2008}. They argue that these sources are the cores of low-luminosity, compact radio galaxies (potentially faint Fanaroff-Riley I (FRI) sources, \citealt{FR1974}), which are not accounted for in models of the faint radio sky. VLBI observations are the ideal tool for probing these compact AGN, which are poorly understood. Furthermore, \citet{Hale2023} derive MeerKAT source counts at $\sim15-100~\mu$Jy flux densities that are larger than model predictions; however, the relative split between AGN and star formation remains unclear. The role of low luminosity AGN and their compact radio emission properties is important to discern to understand its role in galaxy evolution, and larger VLBI-detected samples at low flux densities clearly provide a unique perspective \citep[e.g.][]{Herrera2016}. Improved statistical power of VLBI source counts at the $ S_{\rm 1.4GHz}\lesssim 100 \ \mu$Jy level over a wider range of extragalactic legacy fields will provide unique insights and constraints on the relative distribution of AGN and star-formation powered radio sources at low radio luminosity. \\
\\
\noindent{\bf Host galaxy morphologies of VLBI-selected AGN: }
The CANDELS programme is an {\sl HST} near-infrared through ultraviolet legacy survey \citep{Grogin2011,Koekemoer2011} that studied the evolution of black holes and galaxies between $z = 1.5-8$ and revealed a number of host galaxy properties of X-ray-selected AGN at intermediate to high redshift \citep[e.g.][]{Kocevski2012}. The VLBA central pointing is based on the deeper, high-fidelity imaging of the CANDELS chip positions, enabling detailed analyses of VLBI-selected AGN host morphology and comparison with X-ray selection (and lower resolution radio), thereby probing the question of compact jet-triggering as a function of environment, addressing from a VLBI perspective, seemingly conflicting results on whether or not (1) major mergers play the dominant role in triggering AGN activity at higher redshifts \citep[e.g.][]{Hewlett2017,Marian2019} as outlined in the classical \citet{Sanders1988} scenario, where the merger-induced loss of angular momentum leads to black hole accretion \citep[e.g.][]{Hopkins2006}; and (2) whether AGN hosts at $z \sim 1-3$ are predominantly disk-like in morphology \citep[e.g.][]{Schawinski2011}, and comparison with VLBI-selected local AGN \citep[e.g.][]{Kaviraj2015}. The CANDELS fields have unparalleled optical/infrared image quality that enables the robust morphological modelling of the host galaxies required for these lines of study.\\
\\
\noindent{\bf Probing the population of obscured AGN:}
Despite their significant cosmological importance, obscured quasars remain an elusive population in multi-wavelength surveys. Observations appear to confirm \citet{Silk1998} and \citet{Fabian1999} predictions that the Compton-thick AGN space density increases significantly towards higher redshift \citep{Gilli2001,LaFranca2005,Hopkins2006,Treister2009,Ueda2014,Gilli2022}. While sophisticated selection techniques to select obscured quasars exist \citep[e.g.][]{MartinezSansigre2005}, the dust insensitive, high-brightness temperature filter that VLBI observations provide makes this an important, complementary contribution towards obscured AGN identification. This is supported by \citet{Delvecchio2017}, \citet{Radcliffe2021}, and \citet{Whittam2022}, who all show evidence that no single classification technique can reliably identify all VLBI sources in extragalactic fields as AGN, making these important detections to understand with greater statistical power. \\
\\
\noindent{\bf The search for binary/dual and recoiling AGN: } 
From a theoretical standpoint, we expect binary supermassive black holes to be common in the Universe \citep{Begelman1980,Colpi2011}. However, our observations at present do not agree with this forecasted ubiquity \citep{Burke-Spolaor2011,Koss2012,Comerford2013,Colpi2014,Deane2015,DeRosa2019}. This is a crucial disparity to reconcile as dual/binary supermassive black holes are predicted to play a significant role in galaxy evolution \citep[e.g.][]{Merritt2005,VanWassenhove2012,Mayer2013}, for which observations show evidence, although spatial resolution can limit the ability to decouple this from the galaxy merger process in general \citep[e.g.][]{Komossa2003,Comerford2013,Ellison2013}. Furthermore, binary supermassive black holes are expected to dominate the recently detected stochastic gravitational wave background at nanoHz frequencies \citep[e.g.][]{Sesana2008,Shannon2015,Burke-Spolaor2019,Agazie2023GW,Antoniadis2023paperIII,Reardon2023}, with a poorly constrained dampening factor that gaseous environments and orbital eccentricity expected to play though their modification of the binary in-spiral rate \citep[e.g.][]{Ravi2014,Agazie2023bSMBH,Antoniadis2023paperIV}. VLBI has been shown to be an excellent method to discover binary/dual AGN (e.g. even in the classical radio galaxy, Cygnus~A, \citealt{Perley2017}); as well as wide-field surveys \citep[e.g.][]{Herrera2017,Njeri2023}. Offset, potentially recoiling AGN are also expected during the merger process, with several candidates identified, simulations developed, and large-scale searches underway \citep[e.g.][]{Civano2010,Blecha2016,Hwang2020}. The angular resolution of VLBI provides a unique perspective, and further, the environments are expected to be gas-rich and dust-obscured \citep[e.g.][]{Satyapal2017}, adding impetus on high angular resolution radio observations in this multi-wavelength, multi-messenger field of astrophysics. 

\noindent{\bf Serendipitous search}
As outlined earlier, wide-field VLBI has demonstrated its ability to discover rare, astrophysically important objects (e.g. gravitational lenses, binary supermassive black holes; \citealt{Herrera2017,Spingola2019}). As radio surveys increasingly improve the ability to probe the dynamic radio sky towards the SKA era \citep[e.g.][]{Bignall2015,Fender2015,Mooley2016,Radcliffe2019,Sarbadhicary2021}, advances in wide-VLBI approaches offer a unique discovery technique, opening up the milli-arcsecond scale parameter space over increasingly wider areas with higher sensitivity.

\section{Survey Technical Design} \label{sec:surveydesign}

\begin{figure*}
    \centering
    \includegraphics[width=\textwidth]{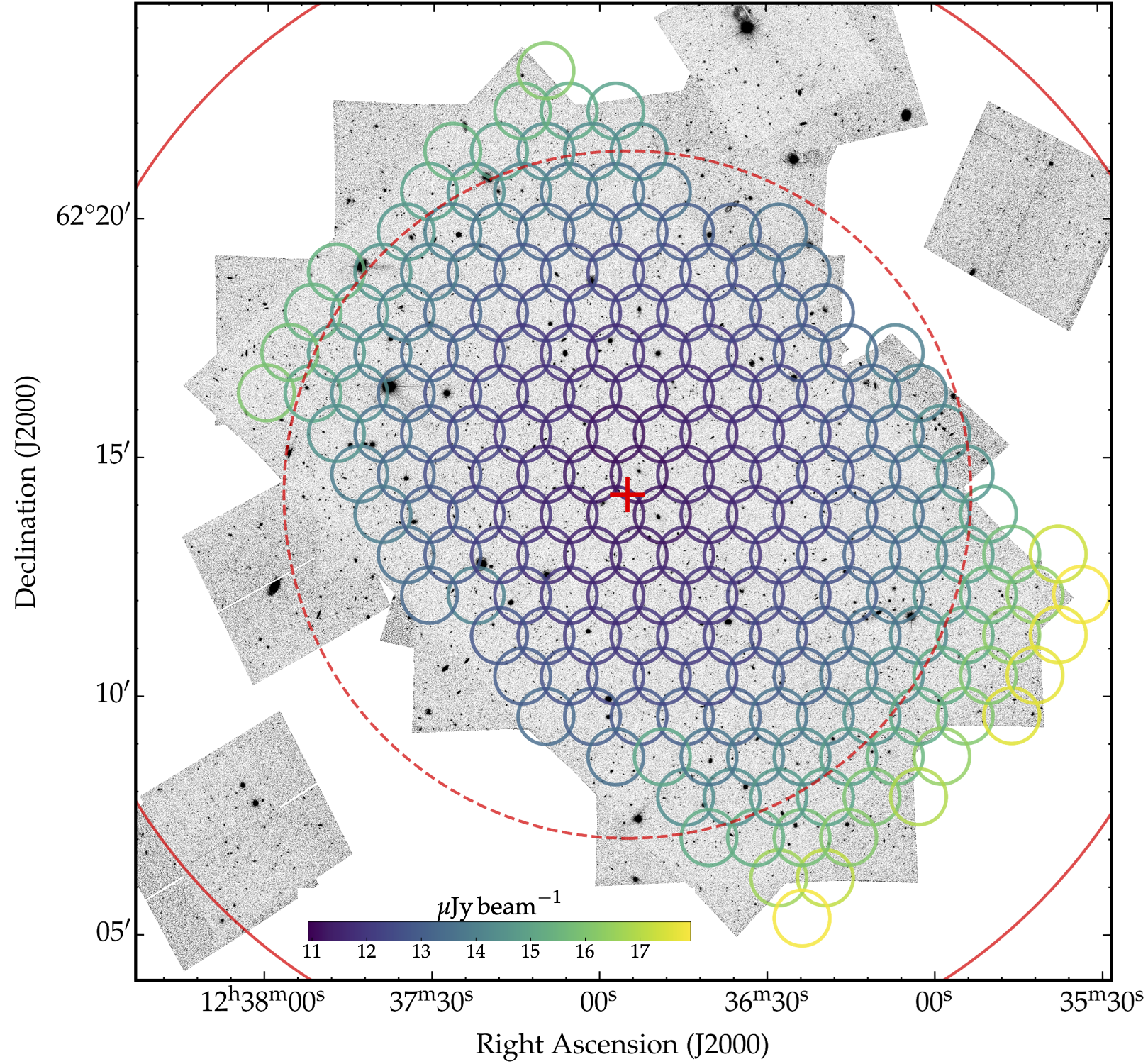} 
    \caption{VLBA phase centres (small diameter colour circles) over-plotted on the {\sl Hubble Space Telescope} F\,606W mosaic of the CANDELS GOODS-North field \citep{Grogin2011,Koekemoer2011}. Each phase centre has a radius of 35~arcsec, the approximate angular distance at which the combined time and bandwidth smearing point source sensitivity loss is at the $\sim$20 per cent level. The red central cross shows the VLBA pointing centre, while the dashed and solid red circles show the 80 and 50 per cent primary beam response contours, respectively. Each phase centre is colourized by the corresponding image's achieved noise rms. }
    \label{fig:PClayout}
\end{figure*}

Several scheduling, processing, and software considerations had to be made in the design of this survey, which we describe in this section. The primary objectives were to cover the CANDELS GOODS-North area of 160\,arcmin$^2$ with contiguous, relatively uniform sensitivity while still retaining practical correlator and archive resource requests, as well as feasible processing requirements to carry out the imaging. A similar approach was taken to generating a large area (77\,arcmin$^2$), quasi-uniform sensitivity map of the nearby face-on spiral galaxy M\,51 to search for supernovae, X-ray binaries, and other transients \citep{Rampadarath2015}. As discussed in Section~\ref{sec:intro}, if contiguous quasi-uniform sensitivity is the objective, then there are two possible strategies: either one must record the data at sufficient time and frequency resolution to image the entire field of interest; or configure multiple phase centres on a regular grid with a spacing based on time and bandwidth smearing considerations. Our selection of the latter approach requires dramatically lower instantaneous computational cost and available random-access memory than the former (particularly for imaging), however, it is still significantly more computationally expensive than the targeted approach of only placing phase centres on known arcsec-scale radio sources and imaging a small (few arcsec$^2$) region at those locations.

At the time of the proposal, our assessment was that image sizes of approximately 64,000$\times$64,000 (64k hereafter) would be near the practical limit of the compute resources we had at our disposal. This imaging consideration set the angular area that would be imaged by each phase centre to approximately $64 \times 64$~arcsec$^2$. To limit time and frequency smearing losses to the $\lesssim$20 per cent level, the correlator dump time and channel width were set to 2 seconds and 250 kHz, respectively. 

The phase centres were positioned to follow a standard hexagonal mosaic pattern used for radio surveys with multiple pointings (see Figure~\ref{fig:PClayout}). However, instead of arranging pointings that lie at the half-power point of their neighbour in Right Ascension to critically sample the sky, we position our phase centres using the constraint that no part of the field should have smearing sensitivity losses larger than the selected $\sim$20 per cent level for $\sim$4000~km baselines. This results in a configuration with a `triangle' of adjacent phase centres that intersect at a radius of $\sim$35~arcsec. Note that phase centres are not jointly imaged since they are simply phase-rotated versions of one another and stem from the same original electric field measurements captured by the telescope, therefore, the noise is not independent. The locations were originally selected based on the planned {\sl HST} chip positions, but as can be seen in Figure~\ref{fig:PClayout}, there is imperfect coverage, depending on which {\sl HST} filter is considered.     

The configuration described required a total of 205 phase centres, which lay within another design constraint; the VLBA correlator limits on the total data output rate, which in turn limits the total number of phase centres for a given observational setup. The total data set size of all 205 phase centres is approximately 4 Terabytes, in {\sc fitsidi} format.

Our approach of a regular grid of phase centres also has the disadvantage of a known source potentially being located in a region with higher noise rms than the full array sensitivity that a co-located phase centre would offer. Another disadvantage, or at least potential additional complexity to the data processing, is the primary beam correction (particularly if imaging is performed near or beyond the half power radius), which is described in Section~\ref{sec:PBcorrection}. Despite these drawbacks, we choose to explore this approach with an emphasis on the possible variable/transient discovery parameter space offered by the quasi-uniform sensitivity, as well as our ignorance of the location of sources that may cross the detection threshold once multi-source self-calibration is applied in future work. Quasi-uniform sensitivity may become especially important in deriving robust source counts with the expected additional detections from multi-source self-calibration. Given the flattening of the source counts near the detection threshold (described in Section~\ref{sec:counts}), these new detections could be numerous and so the quasi-uniform sensitivity is an advantage.

\section{Observations}\label{sec:obs}

The survey observations were carried out over 12 epochs, from September 13, 2013 to November 2, 2013, VLBA project code BD176 (PI: Deane). We use a standard VLBA continuum observing setup with $8\times32$~MHz frequency subbands ranging between 1392 MHz to 1744 MHz (each starting at $\nu$ = 1392.121, 1424.121, 1456.121, 1488.121, 1552.121, 1584.121, 1648.121, 1712.121~MHz) in both parallel circular polarisation hands. The pointing centre is R.A. 12$^\mathrm{h}$36$^\mathrm{m}$55\fs000 and Dec. $+62^{\circ}14\arcmin15\farcs00$ (J2000 coordinates). Data were recorded at bit rate of $1024\,\mathrm{Mbit\,s^{-1}}$ ($8\times32$\,MHz bands, 2-bit sampling, RR and LL polarisations). A correlator dump time of 2\,s and channel width of 250\,kHz was used for the reasons described as part of the survey technical design in Section~\ref{sec:surveydesign}.

\begin{figure}
    \centering
    \includegraphics[width=0.5\textwidth]{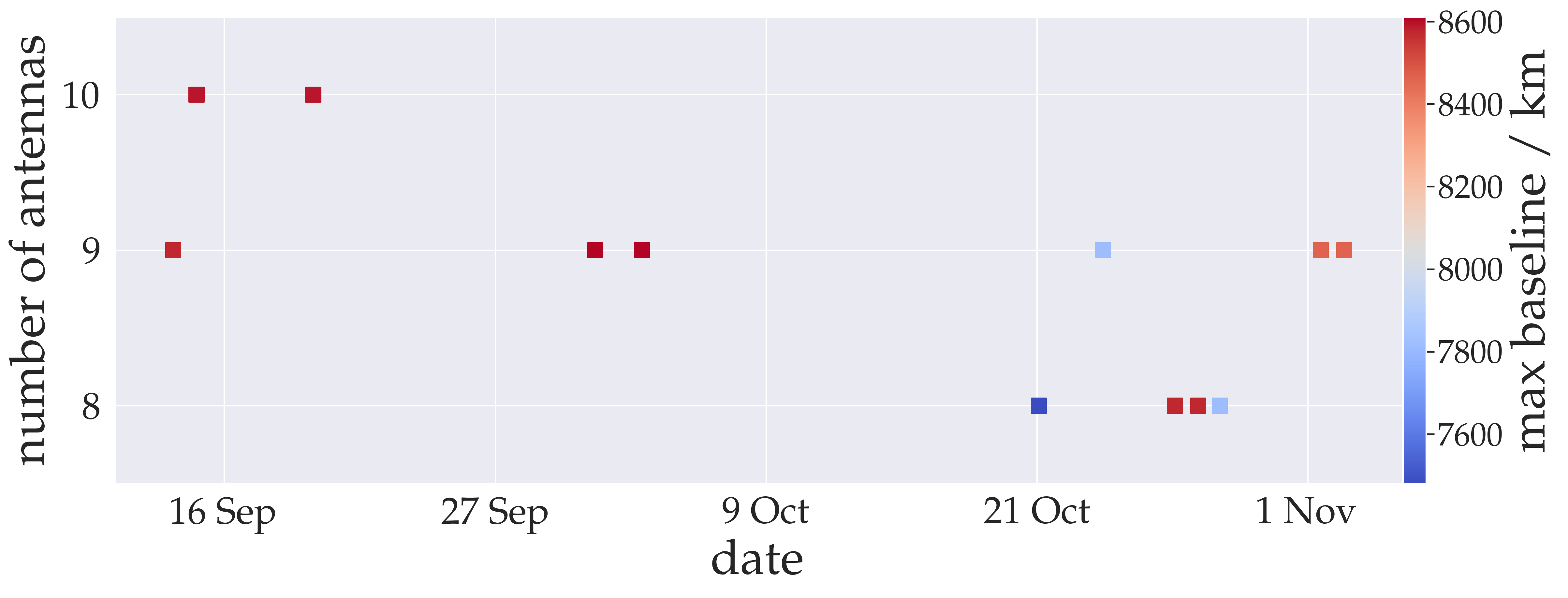}
    \caption{Dates that each of the twelve survey schedule blocks were carried out on the VLBA in the filler mode. The maximum baseline length and number of participating antennas for each schedule block are indicated.}
    \label{fig:obsdates}
\end{figure}

\begin{figure}
    \centering
    \includegraphics[width=0.5\textwidth]{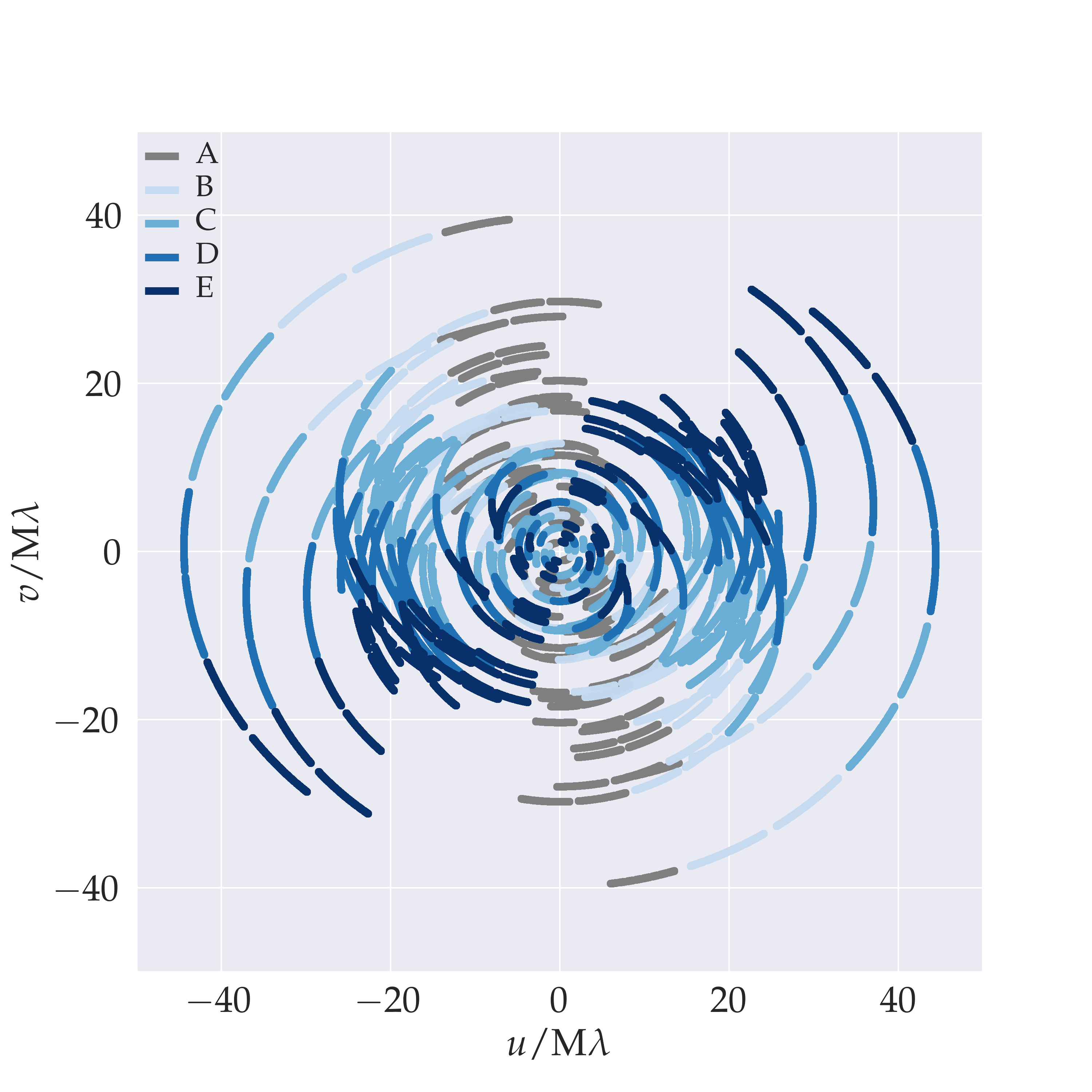}
    \caption{VLBA \emph{uv-} coverage of the GOODS-North field, with colours indicating the five scheduling variations (A-E) that the twelve $\sim$2-hr schedule blocks were split into (see Table~\ref{tab:SBsummary}). }
    \label{fig:uvcoverage}
\end{figure}

A total of 24\,hr of observing time was split into twelve approximately 2\,hr schedule blocks to ease schedulability, detailed in Table~\ref{tab:SBsummary}. Each of these was chosen with suitably chosen starting hour angles in order to improve the combined {\sl uv}-coverage (see Figure~\ref{fig:uvcoverage}), however, not all had the full complement of 10 antennas participating (see Figure~\ref{fig:obsdates}).

For each $\sim$2-hr schedule block, J0927+390, was observed for five minutes as a fringe finder (i.e. solved for delay and delay rate errors). The observations were made using the standard phase referencing mode with J1234+619 used as the complex gain calibrator, which is approximately 24\farcm7 from the target field pointing centre. This calibrator was observed for one minute every five minutes. In total, the on-source integration time on the GOODS-North field was approximately 15.2 hours. This gives an expected thermal noise of $10~\mu\mathrm{Jy\,beam}^{-1}$, assuming that 20 per cent of data was lost to radio frequency interference (RFI).

\begin{table}
	 	\centering
	 	\caption{Summary of the VLBA CANDELS GOODS-North observing epochs.}
	 	\label{tab:SBsummary}
	 	\begin{tabular}{ccc} 
	 		\hline
	 		Epoch ID & Date & Time range (UTC)  \\
	 		\hline
	 		A1 & Oct 23, 2013& 11:04:21--13:02:54 \\
	 		A2 & Oct 28, 2013 & 10:44:41--12:43:14\\
	 		B1 & Oct 26, 2013 & 12:52:14--14:50:47 \\
	 		B2 & Oct 27, 2013 & 12:48:18--14:46:51 \\
	 		C1 & Sep 13, 2013 & 17:40:58--19:39:31 \\
	 		C2 & Sep 14, 2013 & 17:37:03--19:35:36 \\
	 		C3 & Sep 19, 2013 & 17:17:24--19:15:55  \\
	 		D1 & Oct 1, 2013 & 18:34:51--20:33:26 \\
	 		D2 & Oct 20, 2013 & 17:15:10--19:13:45 \\
	 		D3 & Oct 3, 2013 & 18:22:00--20:20:36 \\
	 		E1 & Nov 1, 2013 & 18:27:39--20:26:12 \\
	 		E2 & Nov 2, 2013 & 18:23:43--20:22:16\\
	 		\hline
	 	\end{tabular}
\end{table}

\section{Calibration}\label{sec:cal}

\begin{figure*}
	\includegraphics[width=\textwidth]{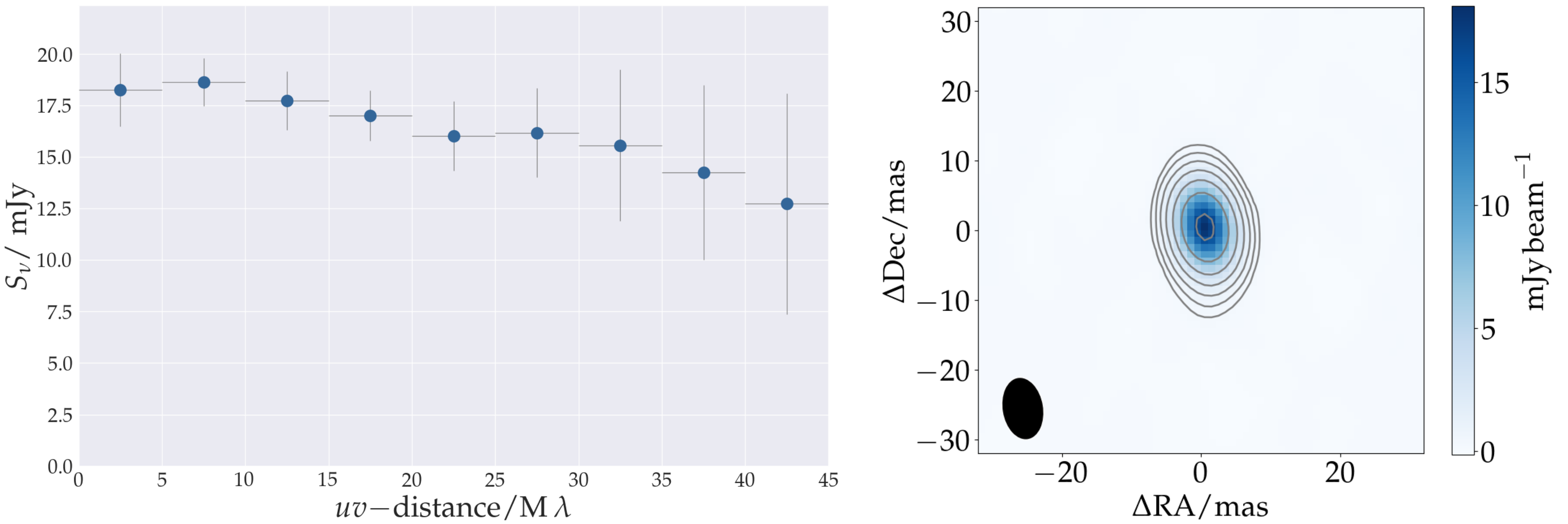}
	\caption{Amplitude versus \emph{uv}-distance (left), and naturally-weighted {\sc Clean} image of J1234+619 (right), the complex gain calibrator used in the VLBA CANDELS GOODS-North survey. J1234+619 has a peak brightness of 17.83\,mJy\,beam$^{-1}$ and an integrated flux density of 18.84 mJy. The source is marginally resolved but still above 10\,mJy level at $>8000$\,km baselines. The contours are drawn at $S_\nu = \pm0.25$\,mJy\,beam$^{-1}$ and increase by factors of two. The map rms is $0.22\,\mathrm{mJy\,beam^{-1}}$. The restoring beam is shown as a black ellipse at the bottom left of the map, with dimensions of $9 \times 5.9$~mas and a position angle of 9.3 deg east of north. }
	\label{fig:phase_cal}
\end{figure*}

Calibration of these data was conducted entirely using the Common Astronomy Software Applications (CASA) software \citep{McMullin2007,CASATeam2022}. This software package now has all the tools necessary to calibrate VLBI data from the raw correlator data to science-ready images \citep{Janssen2019,vanBemmel2019,CASATeam2022,vanBemmel2022}. The multiple phase centre correlation method only includes the calibrator sources in one of the output data sets, not all. Any antenna-based calibration derived for this data set containing all calibrators is applied to all other phase centres to carry out direction-independent calibration. Here, we outline the calibration steps performed, noting the relevant CASA tasks in parentheses. 

The data from each individual epoch was converted from {\sc fitsidi} format to a CASA-compatible measurement set (\textsc{importfitsidi}) and concatenated together so that all of the separate epochs could be calibrated together (\textsc{concat}). Next, a priori calibration was derived. This includes corrections for the errors in the sampler thresholds ({\sc accor}), the conversion of system temperature measurement ($T_\mathrm{sys}$) into a CASA-compatible calibration table - to permit accurate flux scaling ({\sc gencal}) - and the derivation of the VLBA gain curves ({\sc gencal}).

RFI was excised using the {\sc AOFlagger} software \citep{Offringa2012}. Approximately 5~per~cent of the channels at the edges of the spectral windows were removed, and auto-correlations were flagged ({\sc flagdata}). Instrumental delays (the delay induced by differing the electronic paths from receiver to disk/correlator across the bandwidth) for each epoch were derived using a 2-minute solution interval on the bright source J0927+390 (\textsc{fringefit}). The application of these solutions removes phase discontinuities between the spectral windows. The delay rates were set to zero to avoid interpolation errors in time when these solutions are applied. We then derived normalised bandpass corrections using J0927+390 (\textsc{bandpass}). Next, time-variable delays, phase and delay rates were derived for each scan on the complex gain calibrator J1234+619 (\textsc{fringefit}). As the phase jumps between spectral windows were removed earlier, the spectral windows can now be combined to increase the signal-to-noise ratio (SNR) when deriving solutions. A small proportion of solutions ($\sim 5$~per~cent) failed, which we attribute to the phase calibrator being relatively weak ($S_{\rm 1.6GHz} \sim 18\,\mathrm{mJy}$). To minimize the loss of valid data, these flagged data were recovered by linearly interpolating between the nearest good solutions.

These solutions were applied to the phase calibrator source (\textsc{applycal}), and the phase calibrator was imaged (\textsc{tclean}). It was found that the phase calibrator is marginally resolved, as is clearly seen in both the visibility and image domains (see Figure~\ref{fig:phase_cal}), therefore, the delay, rates and phase corrections were refined using a model of the source. These solutions improved the signal-to-noise ratio of the phase calibrator image by $\sim$10~per~cent. Self-calibration was then conducted on the phase calibrator to refine the amplitude and phase solutions. Typical VLBI observations at GHz frequencies will correct for the dispersive delays caused by the ionosphere. However, this correction is currently not available in CASA, so phase self-calibration was performed without combining the spectral windows together. This allowed us to approximate the ionospheric dispersive delays across the bandwidth using a step-wise approximation. Finally, amplitude self-calibration was performed to correct for $T_\mathrm{\mathrm{sys}}$ fluctuations and variable antenna gains over the course of the observations. 

With the direction-independent, antenna-based calibration products derived, these were applied to the target fields in all 205 data sets. It is worth noting that each phase centre data set was individually flagged using \textsc{AOFlagger}, rather than the flags being transferred from a single phase centre. This is due to the different levels of RFI decorrelation at different phase centre coordinates, requiring each to be flagged individually for optimum RFI excision. This means that the flags for each data set are unique and thus need to be flagged individually \citep[J. Morgan / M. Argo private communication;][]{MorganJ2013}.

\subsection{Primary beam correction}\label{sec:PBcorrection}

As demonstrated in \citet{Middelberg2013}, the primary beam of the VLBA is very well approximated as an Airy disk with a diameter, $D=25.48\,\mathrm{m}$. The Airy disk is simply the Fraunhofer diffraction pattern of a uniformly illuminated dish and is given by:

\begin{equation} 
I(\theta)  =I_0\left(\frac{2J_1\left(\frac{\pi}{\lambda}D\sin(\theta)\right)}{\frac{\pi}{\lambda}D\sin(\theta)}\right)^2,
\label{equ:PBresponse}
\end{equation}

\noindent where $J_1(x)$ is the Bessel function of order one, $\lambda$ is the observing wavelength, and $\theta$ is the radial distance from the pointing centre.  The primary beam is normalised to the maximum response, so $I_0 = 1$, and we assume that the primary beam is radially symmetric.

As discussed, in typical wide-field VLBI observations, the phase centres are pre-selected (based on previous low-resolution radio observations) so that sources of interest are located at the centre of each phase centre. The subsequent primary beam corrections derived are only correct for the centre of the phase centre \citep[these are often corrected in the $uv$-plane; e.g. see][]{Middelberg2013,Cao2014,Radcliffe2018}. However, in this case, the phase centres are not pre-selected based on known radio sources; therefore, the sources of interest are unlikely to be located near the phase centre. This means that the primary beam needs to be corrected across the whole image, or rather, wherever a source appears in that image. Since the VLBA is a homogeneous array, this primary beam correction can be conducted in the image plane, in contrast to heterogenous arrays, where the differing primary beam shapes can cause baseline-based amplitude errors, which must be accounted for in the $uv$-plane. In this survey, we calculate the primary beam response for each candidate detection in our catalogue using its location and Equation~\ref{equ:PBresponse}.  We note that this approach does not account for the so-called `beam squint' of the VLBA, caused by the offset between the beams for the two polarisations. Previous wide-field  VLBI surveys have corrected for this using a frequency-dependent, per-antenna, visibility-based based technique \citep[e.g.][]{Middelberg2013,Herrera2017}. We opt not to apply this as the vast majority of our images are within the 80 per cent power point of the primary beam response; the significant added computational expense; as well as the low SNR of the majority of the sources, meaning this correction would likely be sub-dominant in comparison with the statistical uncertainties. 

The achieved primary beam corrected sensitivities are shown in Figure~\ref{fig:PClayout}. Given that our phase centres are relatively near the pointing centre (within the $\sim$80~per cent power point) and our relatively small fractional bandwidth ($\sim$20~per cent), we assume an effective frequency of 1.6~GHz for all phase centres and VLBI detections.

\section{Imaging strategy and source-finding algorithm}\label{sec:imaging}

\begin{figure*}
    \centering
    \includegraphics[width=0.95\textwidth]{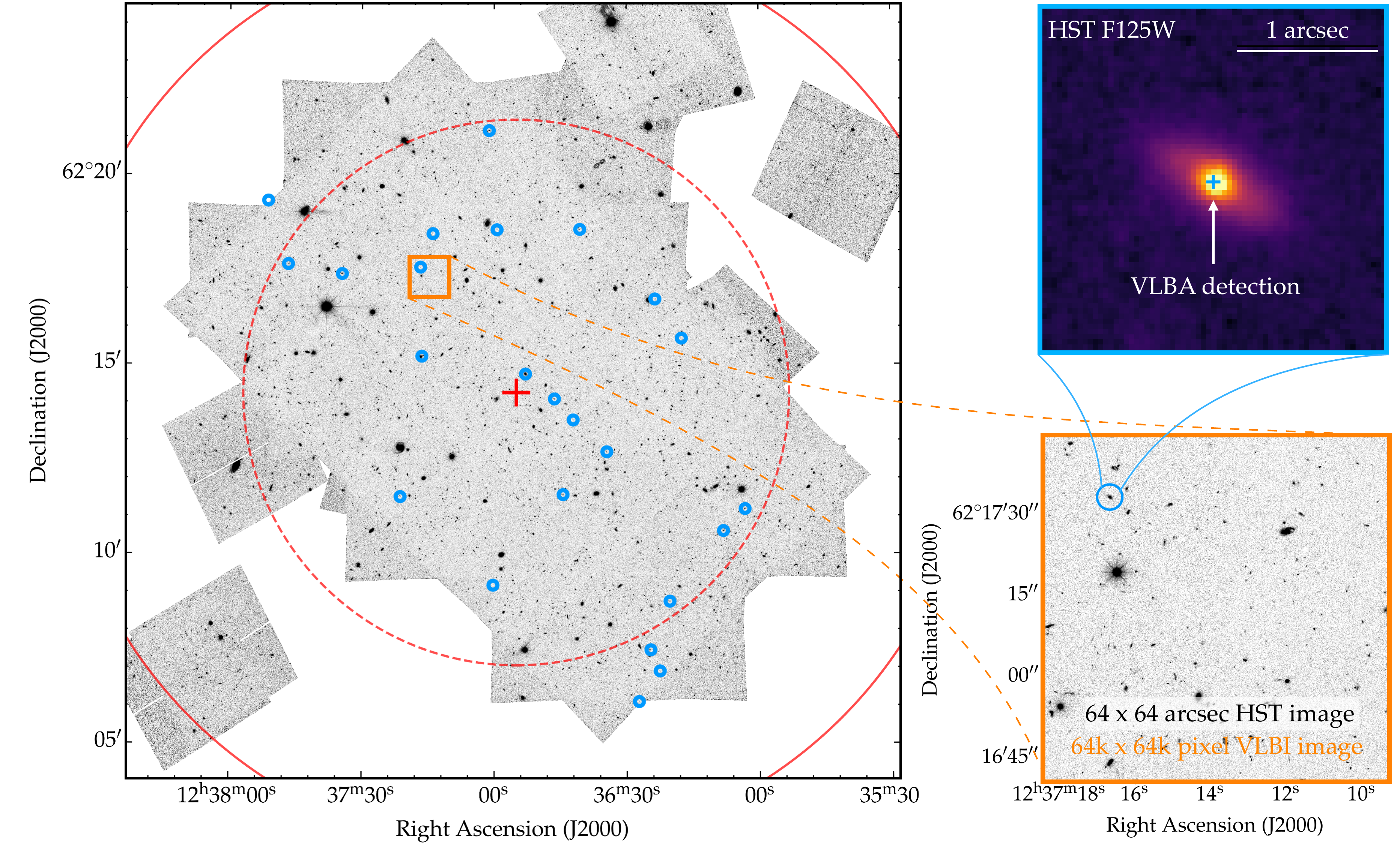}
    \caption{{\bf Left:} Location of 24 VLBA detections presented in the cross-calibration catalogue overplotted on the {\sl HST}\,F1606W map. The red central cross shows the VLBA pointing centre, while the dashed and solid red circles show the 80 and 50 per cent primary beam response contours, respectively. The right panel demonstrates how each of the 205 phase centres is imaged with a size of 64k $\times$ 64k pixels. A further zoom-in is shown to illustrate one of the VLBI detections, which is located within with an {\sl HST}-imaged host galaxy showing a prominent bulge and disk.}
    \label{fig:detectionlocations}
\end{figure*}

As described in Section~\ref{sec:surveydesign}, we choose per-phase centre image dimensions based on several survey design, hardware and software considerations that impact data processing performance. We use {\sc WSClean} to image each phase centre, using a pixel size of 1 milliarcsecond and pixel dimensions of 64,000$\times$64,000. These 64k images are not deconvolved due to the significant additional computational resources required to do so and the negligible benefit this provides for our field. There are no sources that are sufficiently bright to significantly impact the image dynamic range beyond a few arcsec, and therefore, the search for candidate sources. 

For computing efficiency and calculation of the local rms, the 64k images are subdivided into $64\times64$ sub-images (each $1000 \times 1000$ pixels), referred to as 1k sub-images hereafter. We compute the maximum SNR in each 1k sub-image, defining the local rms as the standard deviation of the entire sub-image. The maximum SNR is computed as the maximum pixel value divided by the local rms. We do this for the native angular resolution 1k sub-images, as well as three derived images, which are convolutions of the original 1k sub-images with 2D Gaussian kernels with FWHM of 5, 10, and 20 mas. This approach improves the detection probability of extended sources, as is sometimes employed in more traditional source-finders. Ideally, we would rather employ a uv-taper to generate these smoothed images; however, for data processing and hard disk storage practicalities, we employ the image-domain smoothing approach. 

We employ a stratified limiting SNR threshold to identify candidate sources. In the first, simplest case, we include in our catalogue all VLBI peaks with SNR $> 7\sigma$ in any one of the original or smoothed 1k sub-images. For VLBI peaks with $5.5\sigma < \mathrm{SNR} < 7\sigma$, we only include those with a multi-wavelength source within 0.5 arcsec of the VLBI peak. This multi-wavelength cross-matching is performed using source positions from {\sl Chandra} X-ray \citep{Alexander2003}; {\sl Spitzer} infrared \citep{Ashby2015}, 3D-{\sl HST} \citep{Skelton2014} and VLA 1.4 GHz \citep{Morrison2010}.

In total, we find 24 candidate sources using these selection criteria, the majority of which are selected via the multi-wavelength cross-matching technique. These sources are distributed throughout the survey footprint as seen in Figure~\ref{fig:detectionlocations}. As is detailed in Paper II (Njeri et al., in press), each of these 24 sources are common with catalogues from the EVN, VLA, and e-MERLIN, in addition to multi-wavelength counterparts.

\begin{figure*}
    \centering
    \includegraphics[width=0.82\textwidth]{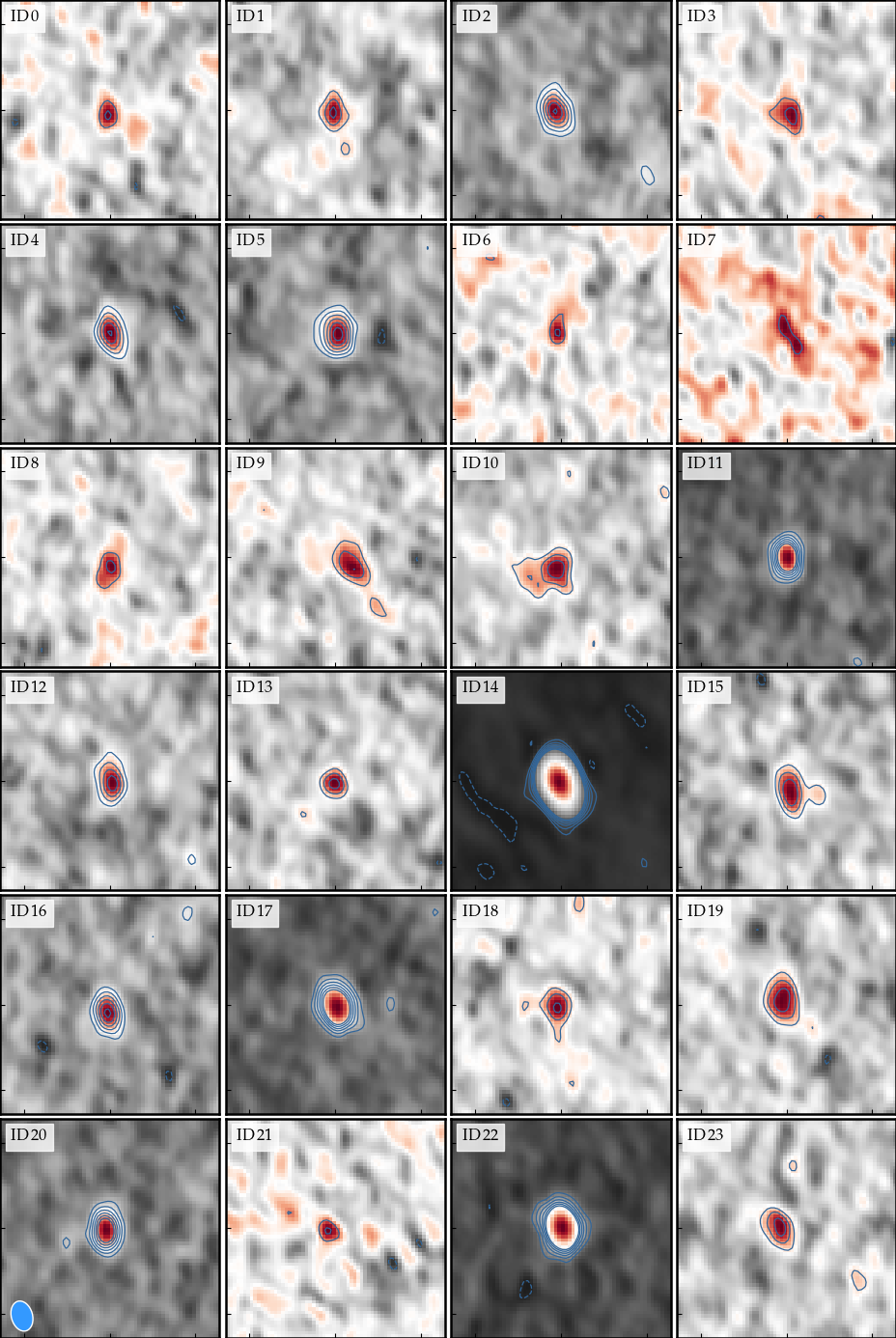}
    \caption{Montage of the 24 detections that make up the cross-calibration catalogue. Each image has an extent of $64 \times 64$~mas$^2$. The contours are drawn at the local rms of the map multiplied by factors of [-3,3,5,7,9,11,13]. See Table~\ref{tab:cat} for source-specific values and coordinates. A representative synthesized beam geometry is shown in the bottom-left panel in blue, which has dimensions $9 \times 6$~mas at a position angle of 9.3 deg. Note that some of the spatially resolved sources only rise above our defined detection threshold when smoothed with a Gaussian kernel; however, they are shown at their native (natural weighting) angular resolution here. See Section~\ref{sec:imaging} for further details. }
    \label{fig:detectionsgrid}
\end{figure*}

Since several sources are spatially resolved, with morphological features of interest, we deconvolve sources once they have been identified in the 64k images, using much smaller $128\times128$ pixel images that are centred on candidate detections using on-the-fly phase rotation. These are {\sc Clean}ed to a depth of $\sim1\sigma$ of the noise level, using both natural and uniform weighting schemes, using a circular mask centred on the peak and with a radius of twice the beam FWHM, down to a noise threshold of $4\sigma$. Two-dimensional Gaussians are used to model the emission using {\sc Casa}'s {\tt imfit} task. These deconvolved images are shown in Figure~\ref{fig:detectionsgrid}, and the resultant catalogue in Table~\ref{tab:cat} lists the integrated flux density from this procedure using the naturally weighted images. Note that because these are deconvolved, the SNR is marginally enhanced compared to the dirty 64k images within which these candidates are first identified. Similarly, they are unsmoothed, so they may also appear below the SNR threshold, which is crossed if convolved with one of the three Gaussian kernels used within the source-finding procedure described earlier. Our aim here is to provide a repeatable method of candidate source selection for more detailed scrutiny using multi-wavelength data and future multi-source self-calibration results. An analysis of the astrometric registration accuracy of the VLBA detections is carried out in Paper II, which finds the astrometry to be consistent with previous \citet{Chi2013} and \citet{Radcliffe2018} EVN 1.6 GHz results within the uncertainties.

\subsection{Statistical Considerations}

In total, this survey generates approximately 0.5 Terapixels of imaging. Therefore, we need to pay careful attention to avoiding spurious detections, given the large number of pixels. As outlined in \citet{MorganJ2013}, under the assumption of a Gaussian noise distribution, with an approximately constant noise rms across the image area considered, the cumulative probability distribution function of the image pixel brightness values is described by, 

\begin{equation}
    \label{equ:cdfdist}
    p(s,\sigma) = \frac{1}{2}\left[ 1 - \erf \left( \frac{s}{\sqrt{2 \sigma^2}} \right) \right],
\end{equation}

\noindent where $s$ is the pixel brightness value and $\sigma$ is the noise rms. Equation~\ref{equ:cdfdist} and our Gaussianity assumptions imply that our imposed $7\sigma$ SNR threshold would result in a total of $N_{\rm spur}\lesssim 0.01$ false-positive VLBI peaks in our maps, given the $\sim 1 \times 10^{10}$ independent VLBI resolution elements (defined by restoring beam dimensions) in what is effectively a 0.5 Terapixel image. This low rate would, therefore, also allow for deviations from a Gaussian noise distribution, for which we find no evidence in our analysis of the wide-field images. Therefore, we are confident that all VLBI peaks above the $7\sigma$ threshold are bona fide sources, and indeed, each of them has a clear multi-wavelength counterpart, just as the $5.5-7\sigma$ detections do (since this is a requirement for their inclusion). 

\begin{figure}
    \centering
    \includegraphics[width=0.48\textwidth]{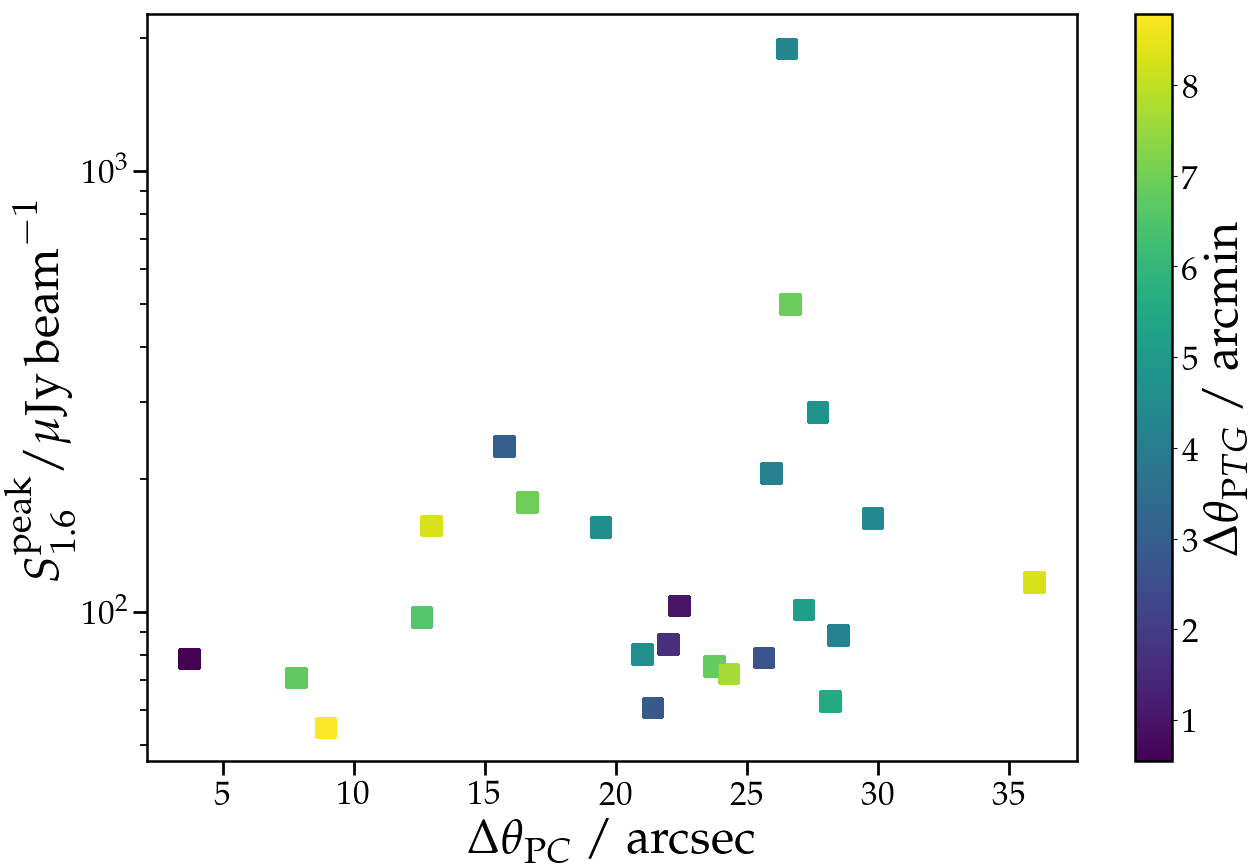}
    \caption{Maximum brightness, $S_{1.6}^\mathrm{peak}$, of each source as a function of the distance from the phase centre, $\Delta \theta_\mathrm{PC}$, of the image within which the source was identified. The colours show the angular distance between each source and the VLBA pointing centre, $\Delta \theta_\mathrm{PTG}$. While this may be in the small number statistics regime, there is no indication of biases in the maximum brightness measured based on source location relative to the phase centre or pointing centre. }
    \label{fig:seps_diagnostics}
\end{figure}

Examining the measured brightness of all 24 detections as a function of their distance from both the nearest phase centres and the VLBA pointing centres reveals no obvious trends (Figure~\ref{fig:seps_diagnostics}), providing some qualitative support that there are no strong biases immediately apparent due to the survey strategy and its practical implementation. This will be further tested with larger source counts in upcoming, wider-field, higher-sensitivity VLBI surveys. 

The false-positive rate is considerably higher for the $5.5\sigma$ threshold, which we expect to result in $N_{\rm spur} \sim94$ spurious peaks within our entire imaging area (i.e. $\sim$0.5 per 64k image).  Attempts to lower the threshold significantly below $7\sigma$ must incorporate additional information to remove false positives, which we do using multi-wavelength catalogues, as previously described. Here, the probability that a $s >5.5\sigma$ noise peak lies within 0.5 arcsec of a catalogued multi-wavelength source is assumed to be,

\begin{equation}
p_{\rm multi} \approx \frac{ N_{\rm spur} n_{\rm s} \pi r_{\rm cross}^2}{\Omega}, 
\end{equation}

\noindent where $n_{\rm s}$ is the cross-matched multi-wavelength catalogue source density, $r_{\rm cross}$ is the cross matching radius, and $\Omega$ is the VLBI imaging area. The source densities for the {\sl Spitzer} infrared, {\sl Chandra} X-ray, {\sl HST} optical/near-infrared, and VLA surveys range from $n_{\rm s} \sim 10^{3-6}$ deg$^{-2}$, meaning that even for the highest source density catalogues used, the probability of cross-matching a VLBI noise peak above $5.5\sigma$ with a multi-wavelength source over the 160 arcmin$^2$ area is less than 2 per cent, a probability comparable to the $7\sigma$ `VLBI-only' threshold, which we deem acceptable for generating a robust cross-calibration in this first paper of the series. We do not take the image-plane smoothing into account when computing the above statistics, however, we do not see any evidence that this leads to spurious detections, as detailed in Paper II.

Naturally, one could argue that the $7\sigma$ `VLBI-only' and $5.5\sigma$ `multi-wavelength' thresholds applied here are somewhat arbitrary, apart from the manual investigation of a range of values we performed and the qualitative assessment thereof. This topic is worthy of a detailed systematic study, which is an enormous computational task and is beyond the scope of this survey overview paper. We explore this topic in a future paper in this series, with several motivations, including the use of low-SNR candidate detections to perform multi-source self-calibration. In principle, including additional sources will improve the quality of the self-calibrated gain solutions and, hence, the image sensitivity and fidelity. Two of the questions this future work will explore are (i) the optimal thresholds in this process and (ii) the resultant false-positive rates.

\section{Survey Data Products}\label{sec:dataproducts}

Here, we describe the four primary survey data products and how they are used in companion papers. 

\begin{enumerate}
    \item There are 205 $\times$ 64k total intensity dirty images used in candidate source identification. Each image is approximately 17~GB in size ({\sc FITS} format) and serves as a reference for comparison with the statistical calibration ensembles to be carried out in a future paper. These images will also be used as a comparison for potential future transient source searches in this field.
    \item A catalogue of candidate sources, referred to as the cross-calibration catalogue. This is a master catalogue from which several derivative catalogues are drawn in Paper II, incorporating a range of comparisons with other radio surveys, as well as multi-wavelength comparisons (Njeri et al., in press). The source-finding approach used to generate this is described in Section~\ref{sec:imaging}. The full catalogue is listed in Table~\ref{tab:cat}, which includes the apparent and primary-beam corrected integrated flux density. All multi-wavelength cross-matching and intrinsic source parameter descriptions are detailed in Paper II.   
    \item {\sc Clean}ed narrow-field total intensity images of each candidate source at a range of {\sc robust} values, with the primary beam correction applied, following the method described in Section~\ref{sec:imaging}. These can be used for more detailed individual analysis of each source, its location within and the morphology of its host galaxy.  
    \item To carry out a transient/variability search, we generated a 64k image for each of the 12 observing epochs, resulting in $12 \times 205 = 2460$ total intensity dirty single-epoch images, each with the same 64k dimensions and a typical rms of $\sigma_{\rm epoch} \sim 38~\mu$Jy\,beam$^{-1}$. This computationally intensive task required the use of {\sc WSClean}'s Image Domain Gridder (IDG) in combination with an {\sc NVIDA} A40 Graphics Processing Unit (GPU) to increase the processing speed. The analysis of these will be reported in a future paper.
\end{enumerate}

\begin{table*}
   \scriptsize
	 	\caption{VLBA CANDELS GOODS-North cross-calibration catalogue. Values are derived from the naturally-weighted images. }
	 	\label{tab:cat}
	 	\begin{tabular}{llllccccccccc} 
	 		\hline 
	 		Source ID & Source Name & RA & Dec & $S_{\rm peak}$ & $\sigma_{\rm peak}$ & $S_{\rm int}$ & $\sigma_{\rm fit}$  & $\Delta$PC$^\dagger$ & $\Delta$Pointing$^{\dagger\dagger}$ & $f_{\rm PB}^\ddagger$  \\
    	 	&   & hms & dms & $\mu$Jy\,b$^{-1}$ & $\mu$Jy\,b$^{-1}$ & $\mu$Jy & $\mu$Jy\,b$^{-1}$ & arcsec & arcmin &  \\
	 		\hline
ID\,0 &  J123603.22+62d1110.61  &  12h36m03.226s  & +62d11m10.6179s & 73 & 11 & 55 & 16 & 7.8 & 6.8 & 0.84 \\ 
ID\,1 &  J123608.12+62d1035.90  &  12h36m08.128s  & +62d10m35.9082s & 80 & 13 & 134 & 33 & 12.6 & 6.6 & 0.84 \\ 
ID\,2 &  J123617.56+62d1540.76  &  12h36m17.564s  & +62d15m40.7679s & 152 & 11 & 210 & 24 & 19.4 & 4.6 & 0.92 \\ 
ID\,3 &  J123620.27+62d0844.26  &  12h36m20.271s  & +62d08m44.2671s & 76 & 12 & 103 & 26 & 23.8 & 6.8 & 0.83 \\ 
ID\,4 &  J123622.50+62d0653.80  &  12h36m22.509s  & +62d06m53.8440s & 167 & 14 & 187 & 28 & 13.0 & 8.3 & 0.76 \\ 
ID\,5 &  J123623.55+62d1642.74  &  12h36m23.553s  & +62d16m42.7437s & 166 & 12 & 248 & 28 & 29.8 & 4.4 & 0.93 \\ 
ID\,6 &  J123624.58+62d0727.28  &  12h36m24.590s  & +62d07m27.2856s & 57 & 13 & 87 & 32 & 24.3 & 7.7 & 0.79 \\ 
ID\,7 &  J123627.20+62d0605.44  &  12h36m27.219s  & +62d06m05.4402s & 41 & 11 & 154 & 52 & 8.9 & 8.8 & 0.74 \\ 
ID\,8 &  J123634.48+62d1240.95  &  12h36m34.484s  & +62d12m40.9582s & 58 & 8 & 82 & 20 & 21.4 & 2.9 & 0.97 \\ 
ID\,9 &  J123640.57+62d1833.00  &  12h36m40.575s  & +62d18m33.0810s & 67 & 10 & 188 & 38 & 21.0 & 4.6 & 0.92 \\ 
ID\,10 &  J123642.09+62d1331.43  &  12h36m42.098s  & +62d13m31.4326s & 75 & 10 & 218 & 39 & 22.0 & 1.7 & 0.99 \\ 
ID\,11 &  J123644.30+62d1133.10  &  12h36m44.395s  & +62d11m33.1710s & 243 & 10 & 252 & 19 & 15.7 & 3.0 & 0.97 \\ 
ID\,12 &  J123646.33+62d1404.60  &  12h36m46.340s  & +62d14m04.6920s & 104 & 9 & 142 & 21 & 22.4 & 1.0 & 1.00 \\ 
ID\,13 &  J123652.89+62d1444.06  &  12h36m52.892s  & +62d14m44.0697s & 81 & 9 & 74 & 15 & 3.7 & 0.5 & 1.00 \\ 
ID\,14 &  J123659.34+62d1832.56  &  12h36m59.342s  & +62d18m32.5666s & 1916 & 31 & 3260 & 79 & 26.5 & 4.3 & 0.93 \\ 
ID\,15 &  J123700.20+62d0909.77  &  12h37m00.255s  & +62d09m09.7779s & 99 & 11 & 171 & 30 & 27.2 & 5.1 & 0.90 \\ 
ID\,16 &  J123701.11+62d2109.62  &  12h37m01.111s  & +62d21m09.6222s & 178 & 13 & 218 & 26 & 16.6 & 6.9 & 0.83 \\ 
ID\,17 &  J123713.87+62d1826.29  &  12h37m13.878s  & +62d18m26.2995s & 286 & 11 & 442 & 27 & 27.7 & 4.7 & 0.92 \\ 
ID\,18 &  J123716.38+62d1512.34  &  12h37m16.382s  & +62d15m12.3441s & 77 & 11 & 108 & 25 & 25.6 & 2.7 & 0.97 \\ 
ID\,19 &  J123716.68+62d1733.31  &  12h37m16.689s  & +62d17m33.3123s & 97 & 10 & 171 & 28 & 28.5 & 4.2 & 0.94 \\ 
ID\,20 &  J123721.26+62d1129.96  &  12h37m21.261s  & +62d11m29.9646s & 209 & 10 & 255 & 20 & 25.9 & 4.1 & 0.94 \\ 
ID\,21 &  J123734.44+62d1722.93  &  12h37m34.445s  & +62d17m22.9329s & 34 & 9 & 78 & 32 & 28.2 & 5.6 & 0.89 \\ 
ID\,22 &  J123746.67+62d1738.59  &  12h37m46.678s  & +62d17m38.5979s & 497 & 15 & 711 & 33 & 26.6 & 6.9 & 0.83 \\ 
ID\,23 &  J123751.24+62d1919.01  &  12h37m51.241s  & +62d19m19.0128s & 122 & 14 & 144 & 28 & 36.0 & 8.3 & 0.76 \\
\hline \\
\end{tabular}
\newline
$^\dagger$ Distance from the nearest phase centre (arcsec).\\
$^{\dagger\dagger}$ Distance from the VLBA pointing centre (arcmin). \\
$^\ddagger$ Primary beam response at the location of the source. 
\end{table*}

\section{Source Counts}\label{sec:counts}

In Section~\ref{sec:imaging}, we present the 24 detections made in this survey over 160~arcmin$^2$. While this is a small sample size that provides relatively poor constraints on the inferred source sky density, the unique feature of this VLBI survey is the quasi-uniform sensitivity over a well-defined area within a well-studied extragalactic legacy field. This is the aspiration goal of future radio facilities that will have baseline lengths ranging from a few tens of metres out to trans-continental scales. In this section, we examine what this first extragalactic survey of its kind is able to contribute to our constraints of mas-scale radio source populations, with a detailed analysis of radio and host galaxy properties presented in Paper II. 

\begin{figure*}
    \centering
    \includegraphics[width=1\linewidth]{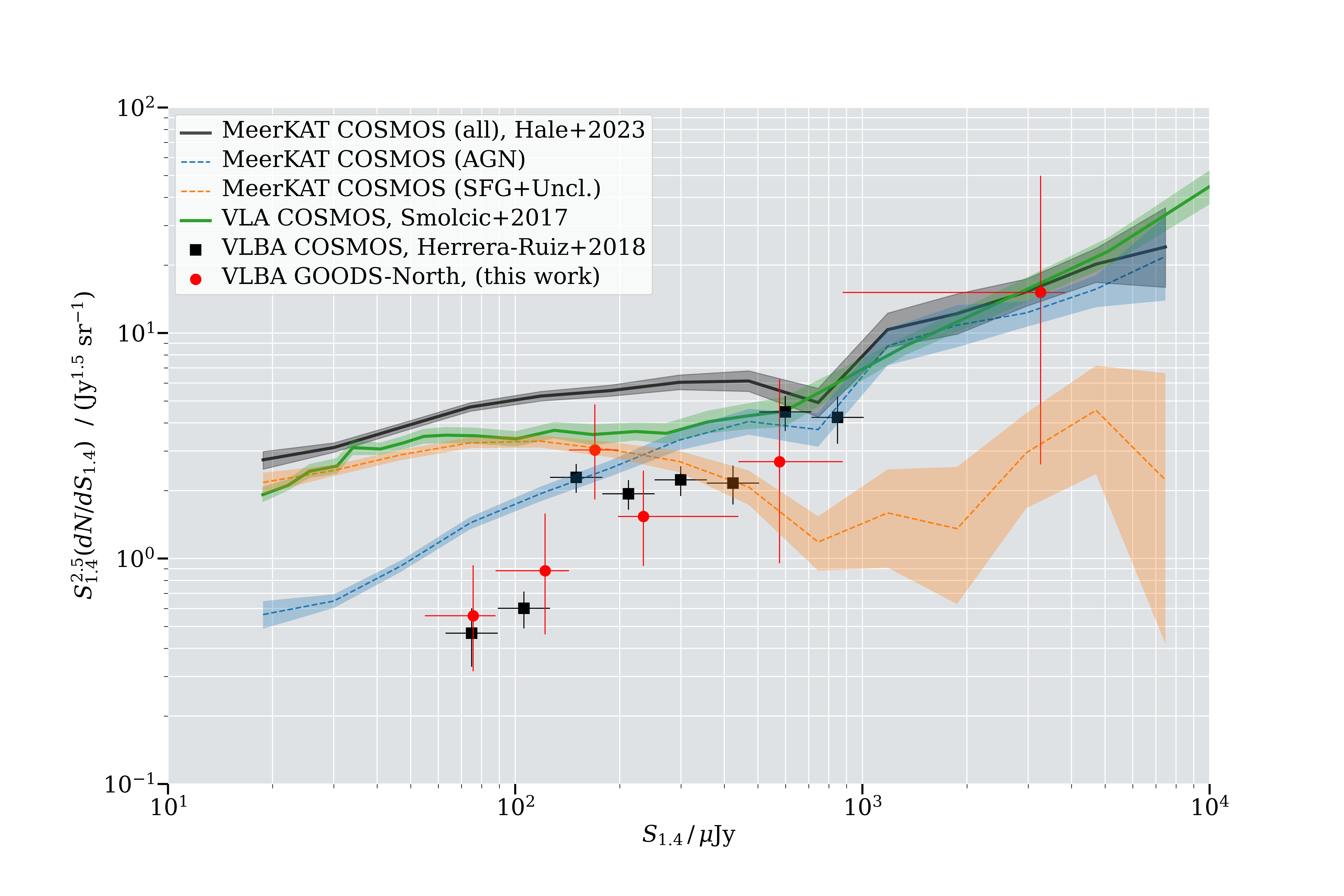}
    \caption{Euclidean-normalized radio source counts for a range of arcsec-scale and VLBI surveys. VLBI source counts are shown as data points, with the VLBA CANDELS GOODS-North Survey (this work) as red circles and VLBA 1.4 GHz COSMOS counts from \citet{Herrera2018} as black squares. The MeerKAT MIGHTEE 1.28 GHz COSMOS total source counts (black solid line) are also shown split into two sub-populations, AGN (blue dashed) and starforming/unclassified (SFG+Unclassified; orange dashed), to assist comparison with the VLBI source counts (see \citealt{Whittam2022} and \citealt{Hale2023} for details on these classifications). The VLA 3 GHz COSMOS \citep{Smolcic2017} counts are shown as a green solid line. All shading envelopes and uncertainties represent 68\,per cent confidence intervals and assume Poisson statistics. All measured flux densities are scaled to 1.4~GHz, assuming a mean spectral index of $\alpha = -0.7$, where $S_\nu \propto \nu^\alpha$.}
    \label{fig:source_counts}
\end{figure*}

Source counts have been used for many decades to better understand the radio sky \citep[e.g.][]{Ryle1958,Condon1984}. Contemporary applications typically show the differential source counts, d$N/$d$S_\nu$, as a function of source flux density, $S_\nu$. At GHz frequencies, this is generally applied to arcsec-scale resolution source counts in total intensity \citep[e.g.][]{Owen2008,deZotti2010,Smolcic2017,Matthews2021} and in polarized intensity \citep[e.g.][]{Hales2014}. Differential source counts were derived by \citet{Herrera2018} using the VLBA COSMOS survey. They argue that the close proximity of the VLBA 1.4\,GHz counts to the VLA 3 GHz source counts \citep{Smolcic2017} was a sign of consistency, implying that most of the lower luminosity radio AGN were accounted for in the VLBA COSMOS survey in the $\sim0.1-1$\,mJy\,beam$^{-1}$ range. In Fig.\,~\ref{fig:source_counts}, we show the derived source counts for the VLBA CANDELS GOODS-North Survey. A point that \citet{Herrera2018} stress is that their Euclidean-normalized VLBI differential source counts are lower limits on the true counts at larger (i.e. arcsec) scales, which do not filter out low-brightness temperature emission. However, this is not strictly correct as the fraction of VLBI-detected sources in a given arcsec-scale radio flux density bin is not constant at all flux densities, as shown in \citet{Deller2014}. So, while it is true that VLBI source counts will typically be lower than arcsec-scale counts at a given flux density, we should be careful not to treat VLBI and arsec-scale differential source counts in a given flux density bin as part of the same population, which the `upper limit' terminology may incorrectly be interpreted as. 

In Figure~\ref{fig:source_counts}, we show the Euclidean-normalized source counts for the COSMOS field for MeerKAT MIGHTEE 1.28 GHz \citep{Jarvis2016,Heywood2022,Hale2023} broken into two sub-populations, as well as the VLA 3 GHz counts \citep{Smolcic2017}, all scaled to 1.4 GHz assuming a spectral index of $\alpha = -0.7$, where $S_\nu \propto \nu^\alpha$. In addition, we show the VLBA (+GBT) mas-scale 1.4 GHz source counts presented in \citet{Herrera2018}, as well as our 1.6 GHz VLBA GOODS-North source counts derived from just 24 detections. While the cosmic variance uncertainty is considerable for an area of 160\,arcmin$^2$ \citep[see][]{Heywood2013}, we are motivated to show this comparison for two primary reasons. First, COSMOS is the extragalactic field with the largest number of VLBI-scale sub-mJy source counts by an order of magnitude. Second, this field includes manual classification of AGN and star-forming galaxies (SFG), allowing direct comparison with these two populations with the compact radio source subsample. 

There are two primary comparisons we wish to highlight from Figure~\ref{fig:source_counts}. The first is the VLBI-only comparison of the VLBA COSMOS and VLBA CANDELS GOODS-North differential source counts. This reveals a consistent profile and similar drop-off at the faint end where completeness is expected to be similar, given this is the same instrument and with comparable angular resolution, observing frequency, and image rms of both surveys (all within $\sim$15 per cent). Second, both sets of VLBI counts follow the AGN-classified profile relatively well, hinting at these two approaches tracing out similar populations. This suggests that a significant fraction of the AGN detected in arcsecond-scale radio surveys in the range $\sim$0.1-1~mJy have a compact core with sufficiently high brightness temperature to be detected with milliarcsecond resolution. This supports the suggestion in \citet{Whittam2017} that the cores of faint radio galaxies are more dominant than previously thought, or at least more dominant than assumed in simulations of the radio source population. 

There are a wide range of potential physical reasons for this observed flattening of the differential source counts of compact radio sources. First, a subset of these may be younger radio jets that have not yet had sufficient time to produce larger extended jets, perhaps still cocooned within the denser ISM of star-forming galaxies. Alternatively, a subset of these sources may belong to a class that fails to produce large-scale jets during their lifetime due to a plethora of reasons, including short duty cycles, weaker jet collimation, and slower jet speed, both of which are potentially linked to lower black hole spins. The latter could be related to the merger history of the respective host galaxies and/or their progenitor supermassive black holes \citep[e.g.][]{Volonteri2005,Reynolds2021}. Further progress on understanding the dominant physical reasons will require large samples enabled by wider surveys at $\lesssim$ 10~$\mu$\,Jy\,beam$^{-1}$ depth. Multi-band VLBI imaging will also likely provide an important additional perspective. 

Figure~\ref{fig:source_counts} demonstrates that while VLBI source counts are still in their infancy, they clearly have the potential to bring a unique perspective to the relative composition of AGN and SFG in the important transition flux density range of $\sim0.1 - 1$\,mJy\,beam$^{-1}$. What is critical to this are well-defined survey areas, as provided in the VLBA CANDELS GOODS-North Survey. Furthermore, we clearly require a wide range of fields with excellent multi-wavelength coverage and preferably with AGN/SFG classification in hand to address cosmic variance and begin to explore the impact of environment as well as the clustering properties of the lower luminosity VLBI-selected radio sources.  This statistical comparison of the source counts between VLBA COSMOS and VLBA CANDELS GOODS-North Survey is a first step, which is followed by a detailed source-by-source comparison of the VLBA, EVN, e-MERLIN, and VLA GOODS-North surveys, which is presented in Paper II of this series, and upcoming, wider-area VLBI surveys at similar depths in other extragalactic legacy fields.

\begin{table*}
	 	\centering
	 	\caption{VLBA CANDELS GOODS-North Survey 1.6 GHz Source Counts, scaled to 1.4 GHz for consistency with the literature (see main text). The lower and upper uncertainties translate to the 68 per cent confidence levels (C.L.).}
	 	\label{tab:srccounts}
	 	\begin{tabular}{cccccccc} 
	 		\hline \hline
	 		 $S_{\rm 1.4, min}$ & $S_{\rm 1.4, max}$ & $\overline{S_{\rm 1.4}}$ & $N$    & $S_{1.4}^{2.5} (dN / dS_{1.4})$ & lower 68\% C.L. & upper  68\% C.L.   \\
                $\mu$Jy & $\mu$Jy &  $\mu$Jy & & Jy$^{1.5}\,$sr$^{-1}$ & Jy$^{1.5}\,$sr$^{-1}$ & Jy$^{1.5}\,$sr$^{-1}$  \\
	 		\hline
55 & 88 & 76 & 5 & 0.56 & -0.24 & +0.38 \\ 
88 & 143 & 122 & 4 & 0.88 & -0.42 & +0.70 \\ 
143 & 198 & 170 & 6 & 3.03 & -1.20 & +1.81 \\ 
198 & 439 & 234 & 6 & 1.54 & -0.61 & +0.92 \\ 
439 & 878 & 577 & 2 & 2.69 & -1.74 & +3.55 \\ 
878 & 3843 & 3260 & 1 & 15.12 & -12.50 & +34.77 \\ 
	 		\hline
	 	\end{tabular}
\end{table*}

\section{Summary}

In this paper, the first in a series, we describe the design, data processing, primary data products, and derived differential source counts of a wide-field, quasi-uniform sensitivity VLBA survey of GOODS-North field. The survey area is 160~arcmin$^2$ and reaches a depth of $\sigma \sim 11 \, \mu$Jy\,beam$^{-1}$ at the pointing centre. The survey serves as a technical demonstration of an alternative approach to wide-field VLBI surveys, placing phase centres on a uniform hexagonal grid rather than on known sources pre-selected by previous arcsec-scale radio or multi-wavelength surveys. We use this approach to generate what would collectively be a $\sim$0.5 Terapixel image, made up of 205 `sub-images', each with 64,000$\times$64,000 pixel dimensions, where each pixel is 1\,mas$^2$. We employ a novel approach to candidate source identification, incorporating smoothing kernels and multi-wavelength cross-matching to derive what we refer to as the cross-calibration catalogue, comprising of 24 sources. The cross-matching and host galaxy analysis is performed in Paper II (Njeri et al., in press), along with a detailed description of the radio properties across a spatial dynamic range of $\gtrsim10^4$. Paper II also performs a comparison of the EVN GOODS-North field \citep{Chi2013,Radcliffe2018} with that of the VLBA, which is the first comparison of its kind for this depth and sample size that the authors are aware of. The survey design also enables a detailed analysis of the performance of statistical calibration \citep[i.e. Multi-Source Self-Calibration,][]{Middelberg2013,Radcliffe2016} and the relevant tradeoffs in the catalogue size versus the lower limit of the flux density, a study detailed in a future paper in this series. The survey was carried out in 12 individual epochs of comparable duration and sensitivity, enabling a transient/variability search through the $205 \times 12$ single-epoch 64k images for sources of interest. 

We derive the VLBI differential source counts for this uniform sensitivity field and show that these are consistent with previous VLBI source counts in the COSMOS field. These broadly trace the AGN population detected in arcsecond-scale radio surveys, with one important deviation: there is a distinct flattening of the source counts in the $\sim$100-500\,$\mu$Jy range. This could suggest a transition in the population of compact radio sources as the host galaxies transition into the starforming population. The physical reasons for the flattening of VLBI source counts are speculative at this point and include the impact of both the denser ISM and lower black hole spin, which could be related to galaxy/black hole merger history.  Increased statistical power and multi-wavelength information will be required to test possible explanations. Multi-band VLBI imaging will likely provide an important additional perspective. 

The scientific and technical demonstrations of this survey may serve as useful inputs to the design and execution of future large-area surveys at milli-arcsecond resolution, including those planned future wide field-of-view African VLBI stations \citep{Gaylard2011,Godfrey2012,Agudo2015,Paragi2015}, which will greatly enhance VLBI access to southern hemisphere extragalactic legacy fields, in concert with the European VLBI Network and Australian Long-Baseline Array, and the Square Kilometre Array mid-frequency array. Furthermore, the technical approach outlined here may be useful in the design of transient surveys and time-critical rapid follow-up VLBI observations or for specific transient classes that can be identified on timescales comparable to the multi-month observational period of this survey.

\section*{Acknowledgements}

We thank the referee, Adam Deller, for a thoughtful and thorough review, which improved the quality of the paper. We would like to thank the NRAO staff, who helped significantly in scheduling these observations. We also thank Ian Heywood and Michael Bietenholz for very useful discussions. RPD and OMS acknowledge funding by the South African Research Chairs Initiative of the DSI/NRF. AAD and JFR acknowledge funding from the South African Radio Astronomy Observatory (SARAO), which is a facility of the National Research Foundation (NRF), an agency of the Department of Science and Innovation (DSI). MJJ and IHW acknowledge generous support from the Hintze Family Charitable Foundation through the Oxford Hintze Centre for Astrophysical Surveys. We acknowledge the use of the ilifu cloud computing facility – www.ilifu.ac.za, a partnership between the University of Cape Town, the University of the Western Cape, the University of Stellenbosch, Sol Plaatje University, the Cape Peninsula University of Technology and the South African Radio Astronomy Observatory. The Ilifu facility is supported by contributions from the Inter-University Institute for Data Intensive Astronomy (IDIA – a partnership between the University of Cape Town, the University of Pretoria, the University of the Western Cape and the South African Radio Astronomy Observatory), the Computational Biology division at UCT and the Data Intensive Research Initiative of South Africa (DIRISA). This project made extensive use of The Cube Analysis and Rendering Tool for Astronomy (CARTA, \citealt{Comrie2021}), and we are grateful to the CARTA developers. This work made use of the Swinburne University of Technology software correlator, developed as part of the Australian Major National Research Facilities Programme and operated under licence. The National Radio Astronomy Observatory is a facility of the National Science Foundation operated under cooperative agreement by Associated Universities, Inc. This work made use of Astropy:\footnote{http://www.astropy.org} a community-developed core Python package and an ecosystem of tools and resources for astronomy \citep{astropy:2013, astropy:2018, astropy:2022}.

\section*{Data Availability}

We expect to make the full set of data products, along with the enhanced or value-added data products, publically available following the publication of the survey paper series. In the interim, the authors may make the data products available upon reasonable request.



\bibliographystyle{mnras}
\bibliography{refs} 


\bsp	
\label{lastpage}
\end{document}